\documentclass[onecolumn,epjc3]{svjour3}
\smartqed  
\RequirePackage{graphicx}
\usepackage{amsmath,amssymb}
\usepackage{color}

\journalname{Eur. Phys. J. C}

\begin{document}

\title{Holographic description of unified early and late universe in viscous mimetic gravity}

\author{G.S. Khadekar\thanksref{e1,addr1}
        \and
        Saibal Ray\thanksref{e2,addr2} 
        \and
        Aritra Sanyal\thanksref{e3,addr3} 
}

\thankstext{e1}{e-mail: gkhadekar@yahoo.com}
\thankstext{e2}{e-mail: saibal.ray@gla.ac.in}
\thankstext{e3}{e-mail: aritrasanyal1@gmail.com}

\institute{Department of Mathematics, Rashtrasant Tukadoji Maharaj Nagpur University, Mahatma Jyotiba Phule Educational Campus, Amravati Road, Nagpur 440033, Maharashtra state, India\label{addr1}
\and
Centre for Cosmology, Astrophysics and Space Science (CCASS), GLA University, Mathura 281406, Uttar Pradesh, India\label{addr2}
\and
A.P.C. College, New Barrakpur, District 24 parganas (N), Kolkata 700131, West Bengal, India\label{addr3}}

\date{Received: date / Accepted: date}

\maketitle

\begin{abstract}
In this study, we explore the mimetic matter model proposed by Chamseddine and Mukhanov [J. High Energy Phys. {\bf 11}, 135 (2013)], utilizing the holographic principle to coherently describe both the early and late universe when bulk viscosity is present in the inhomogeneous equation of state. Our examination of the universe's evolution is based on the generalized infrared-cutoff holographic dark energy model detailed by Nojiri and Odintsov [Eur. Phys. J. C {\bf 77}, 528 (2017)] within the context of the flat FRW model. From a holographic perspective, we derive the energy conservation equation incorporating mimetic matter through a viscous holographic fluid model. Furthermore, we analyze various scenarios of bulk viscosity by assuming a constant equation of state parameter and derive the infrared cut-off expression in terms of the particle horizon. We demonstrate that within the framework of mimetic gravity, there is a class of solutions comparable to those in General Relativity, with an additional contribution from a non-relativistic mimetic matter component. These solutions can effectively describe dark matter.
\end{abstract}

%\keywords{Mimetic gravity; Inhomogeneous equation of state; Bulk viscosity; Dark energy; Dark matter; Holographic viscosity }

%\ccode{PACS Nos.: 98.80 Cq, 89.80Cq., 98.80.$-k$, 98.80Es}

\section{Introduction}

The Holographic Principle, as discussed by Li \cite{Li} and Nojiri et al. \cite{Nojiria}, is a significant theoretical framework for understanding the evolution of the universe. Applying the concept of holographic dark energy can provide insights into quantum gravity. This principle has been applied to both the early and late stages of the universe, as evidenced by the work of Nojiri and Odintsov \cite{Nojirib,Nojiric}, and Elizalde \cite{Elizalde}. Nojiri and Odintsov \cite{Nojirid} proposed a generalized version of cut-off holographic dark energy, which uses a combination of parameters such as the Hubble constant, particle horizon, future horizon, cosmological constant, and the universe's lifetime to determine the infrared cut-off.

In this paper, we explore the concept of mimetic matter, initially proposed by Matsumoto \cite{Jiro} as a hypothesis for dark matter. Later, it was suggested that by including a potential term for the scalar field, this model could also account for dark energy, as shown by Chamseddine and Mukhanov \cite{Chamseddine}. The mimetic matter model is a conformally invariant theory where the physical metric is expressed as a product of an auxiliary metric and the contraction of the auxiliary metric with the kinetic term of the scalar field. This model has been shown to be equivalent to the Lagrange multiplier approach by various studies \cite{Golovnev,Barvinsky}.

The physical metric \( g_{\mu\nu} \) in the Einstein-Hilbert variational principle is represented as follows:
\begin{equation}
\label{eq1}
g_{\mu\nu} = \bar{g}_{\mu\nu} g^{\alpha \beta}(\partial^\alpha \phi \partial_\beta \phi),
\end{equation}
where \( \bar{g}_{\mu\nu} \) is an auxiliary metric and \( \phi \) is a scalar field with the usual dimension \((-1)\) \cite{Brevika}.

In standard cosmology, the $\Lambda$CDM model correlates dark energy with a cosmological constant $\Lambda$, which typically represents energy with negative pressure and an equation of state parameter less than \(-\frac{1}{3}\). Dark matter is generally depicted as non-relativistic matter, specifically cold dark matter (CDM) \cite{Jiro}.

Mimetic gravity is a compelling theory of gravity that describes dark matter and dark energy as geometric effects without introducing new fields of matter \cite{Nashed}. This theory is supported by various observational facts, such as the existence of dark energy and dark matter, which play crucial roles in the universe's evolution \cite{Seljak}. Dark matter provides the necessary gravitational pull for the rotation of galaxies and the formation of large-scale structures in the universe \cite{Nashed}.

Dark energy and dark matter remain two primary research areas in modern cosmology. Despite numerous efforts, dark matter continues to elude direct detection beyond gravitational interactions at galactic scales. Observations such as Cosmic Microwave Background Radiation (CMB) \cite{Komatsu,Adea}, Baryon Acoustic Oscillations (BAO) \cite{Percival,Delubac}, and the large-scale structure of the universe strongly support the existence of dark matter and dark energy. However, the problems associated with these components in the conventional cosmological model remain unresolved. There are many candidates for dark energy and dark matter, but the debate on the most viable model continues. A small positive cosmological constant $\Lambda$ can effectively parameterize dark energy, yet this approach leads to challenges such as the cosmological constant problem and the coincidence problem \cite{Yevgeniya}.

Baffou et al. \cite{Baff} examined the inflationary scenario in viscous $f(R,T)$ gravity with a mimetic potential and a Lagrange multiplier, considering the presence of dark energy alongside a perfect fluid and a bulk viscous fluid. They derived the Friedmann equations that describe cosmological evolution in viscous mimetic $f(R,T)$ gravity.

Brevik and Timoshkin \cite{Brevikb} expanded an axion $F(R)$ gravity model, incorporating bulk viscosity in the general equation of state (EOS) and utilizing the holographic principle to describe the early and late universe coherently. They derived analytical formulas for the infrared cut-offs in terms of the particle horizon, considering various forms of bulk viscosity. Jiro Matsumoto unified the descriptions of dark matter and dark energy in the mimetic matter model, explaining the universe's time evolution and matter density perturbation similarly to the $\Lambda$CDM model \cite{Jiroa}.

Matsumoto et al. \cite{Jiro} further investigated matter density perturbations in the mimetic matter model, demonstrating that mimetic dark matter behaves like CDM at the perturbation level. Elkhateeb \cite{Elkhateeb} studied a dissipative unified dark fluid model, where the dissipative part is a bulk viscosity with a constant coefficient. Brevik and Timoshkin \cite{Brevika} also explored the universe's evolution in a homogeneous and isotropic FRW metric when the EOS has two power-law asymptotes.

Inspired by these studies, this paper characterizes the early and late universe in a mimetic cosmology using a viscous holographic fluid model.

The paper is structured as follows: Sect. 2 introduces mathematical background of (i) mimetic gravity and derives the Friedmann equations for the FRW model in mimetic cosmology and (ii) holographic principle, describing both the early and late-time universe in the context of mimetic dark matter. Sect. 3 discusses the dissipative unified dark fluid model, the structure of the proposed EOS and also applies various thermodynamic parameters and forms of bulk viscosity. The conclusions are presented in Sect. 4.

\section{A brief overview of the mathematical background}

\subsection{Mimetic gravity and cosmology}

The current state of mimetic gravity is briefly reviewed in this section. As per the mimetic gravity theory derived from Equation (1), the physical metric $g_{\mu\nu}$ remains invariant under the conformal transformation \cite{Chamseddinea} of the auxiliary metric $\bar g_{\mu\nu}$. Specifically, under $\bar g_{\mu\nu}$ $\rightarrow$ $\Omega^2$ $\bar g_{\mu\nu}$, we obtain $g_{\mu\nu}$ $\rightarrow$ $\bar g_{\mu\nu}$.

The action of the mimetic gravity in the presence of potential for the mimetic field is given by~\cite{Bita,Ahmad} 
\begin{equation}
\label{eq2}  I = \int d^4 x \sqrt(-g) \left(\frac{R}{2 \kappa^{2}}- \frac{\lambda}{2}(\partial^\mu \phi \partial_\nu \phi-1) + V(\phi) + L_{m} \right)dx,
\end{equation}
where $g_{\mu\nu} = (\bar g_{\mu\nu}, \phi)$, \; $\kappa^2 = 8\pi G$, where $G$ is the Newtonian gravitational constant, $ R $ is the Ricci scalar, $L_{m}$ is the Lagrangian for matter, $V(\phi)$ is the self-interacting scalar field, and $\lambda$ is the Lagrange multiplier. The factor $\frac{1}{2}$ is in front of the  Lagrange multiplier $\lambda$ is introduce for later convenience and $g$ is the determinant of the physical metric. 

Without the requirement for either an explicit dark matter or dark fluid, it has been demonstrated that such a model offers an affordable means of generating a variety of straightforward and well-motivated cosmological scenarios, relevant to both early and late time cosmology \cite{Chamseddine}. The field equations can be obtained by modifying the action Eq. (2) above with respect to the physical metric $g_{\mu\nu}$
\begin{equation}
\label{eq3}  \frac{G_{\mu\nu}}{\kappa^{2}} = \lambda \partial_{\mu} \phi \partial_{\nu} \phi + g_{\mu\nu} V(\phi) + T_{\mu\nu}, 
\end{equation}
where $G_{\mu\nu}$ is the Einstein tensor and $T_{\mu\nu}$ is the energy momentum tensor of the usual matter. 

The variation of the action Eq. (2) with respect to the Lagrange multiplier $\lambda$ gives 
\begin{equation}
\label{eq4} g^{\mu\nu} \partial_{\mu} \phi\partial_{\nu}\phi = 1.
\end{equation}

This constraint which is consistent with conformal transformation Eq. (1). Taking the covariant derivative of the Eq. (2) and using the fact that $\nabla^{\mu}G_{\mu\nu} =0$, together with the continuity equation $\nabla^{\mu}T_{\mu\nu} =0$, we find the Lagrange multiplier to be 
\begin{equation}
\label{eq5} \lambda = \frac{G}{\kappa^{2}}-T-4 V(\phi), 
\end{equation}
where $G$ and $T$ are the trace of the Einstein tensor and matter energy momentum tensor respectively. 

Finally variation of the action Eq. (2) with respect to the mimetic field $\phi$ gives 
\begin{equation}
\label{eq6} \nabla^{\mu}(\lambda \partial_{\mu} \phi) + \frac{dV}{d\phi} =0.
\end{equation}

We would like to apply the mimetic gravity to cosmological setup. We assume the holographic isotropic universe in which its matter content in in the form of perfect fluid, $ T_{\mu\nu} = (\rho +p)u_{\mu}u_{\nu}+ p g_{\mu\nu}$, with $u_{\mu}, \rho,$ and $p$ are respectively the fluid velocity $(u^{\mu}u_{\mu}=-1)$, energy density and pressure of a perfect fluid. A simple calculation give $T^{\mu}_{\nu} = diag(\rho, -p, -p, -p)$. 

We consider homogeneous and isotropic spatially flat FRW line element
\begin{equation}
\label{eq7}  ds^{2}= dt^{2}-a^{2}(t)(dr^{2}+r^{2} d\theta^{2} + r^{2} sin^{2}\theta d\phi^{2}),  
\end{equation}
where $a(t)$ is a scale factor of the universe.

For homogeneous cosmology, the mimetic scalar field is a function of time, i.e. $\phi = \phi(t)$. Thus from the Eq. (4) we have $\dot\phi^2=1$ and $\phi =t$, where we have also set constant of integration equal to zero.

The cosmological field equations can be derived by substituting the metric Eq.  (7) and also the energy momentum tensor in the field Eqs. (3) and (6), we can find 
\begin{equation}
\label{eq8} 3 H^2  = \kappa^2( \rho +\lambda + V),
\end{equation}

\begin{equation}
\label{eq9} 2\dot {H} + 3 H^2 = -\kappa^2 (p-V),   
\end{equation}

\begin{equation}
\label{eq10} \dot \lambda + 3H\lambda + \frac{dV(\phi)}{d\phi}  = 0,  
\end{equation}
where $H=\frac{\dot a}{a}$ is the Hubble parameter. Note that since $\phi = t$, thus $ V(\phi) = V(t)$ is only a function of $t$ similar to the $\lambda(t)$, $\rho(t)$ and $p(t)$.

Using $\phi(t) = t$, Eq. (10) can be express as 
\begin{equation}
\label{eq11} \dot \lambda + 3H\lambda + \frac{dV(t)}{dt}  = 0.
\end{equation}

Thus if we assume
\begin{equation} 
\label{eq12}V(t) = -\frac{\beta}{(1+\beta)} \lambda(t). 
\end{equation}

Then after solving Eq. (11) analytically to get 
\begin{equation} 
\label{eq13} \lambda =  \lambda_{0} a^{-3(1+\beta)}, 
\end{equation}
where $\beta$ is dimensional constant and $\lambda_{0}$ is the present value of $\lambda$ and we choose $a_{0} = a( t
=t_{0}) =1$. 

Hence the explicit form of mimetic potential is given by 
\begin{equation} 
\label{eq14}V(a) = -\frac{\beta}{(1+\beta)} a^{-3(1+\beta)}. 
\end{equation}

In the absence of the mimetic potential $V$, i.e. ($\beta= 0 = V)$, then Eqs. (8) and (9) are the usual cosmological equations with a new non-relativistic matter component given by the Lagrange multiplier $\lambda$, which can indeed be used to model dark matter \cite{Jibitesh}. Therefore, from Eq. (13) we can easily obtain 
\begin{equation} 
\label{eq15} \lambda =  \lambda_{0} a^{-3}. 
\end{equation}

So, basically, the above equation mimics dark matter~\cite{Chamseddinea}. In the present model we assume that the mimetic potential is a function of $\lambda$, which plays the role of dark matter.

\subsection{Holographic description of universe}

The essential features of the holographic principle are presented in this section, using the Ref.~\cite{Li}  to indicate the relationship between the infrared-cutoff and the short distance cut-off. According to the holographic principle, all of the physical quantities in the universe, including the dark energy density, can be explained by fixing certain quantities at the boundaries of the universe~\cite{Hawking}.

The energy density is inversely related to the squared infrared cut-off, according to the holographic principle, so that $L_{IR}$~\cite{Li}
\begin{equation}
\label{eq16} \rho = \frac{3c^2}{\kappa^2 L^2_{IR}},
\end{equation}
where $c>0$ is a non-dimensional positive constant.

There are several ways for choosing the infrared-cutoff, identifying the particle horizon $L_{p}$ or alternatively with future horizon $L_{f}$~\cite{Nojirib}
\begin{equation} 
\label{eq17} L_{p}  = a(t) \int_{0}^{t}\frac{dt'}{a(t')}, \;\;\;  L_{f}  = a(t) \int_{t}^{\infty}\frac{dt'}{a(t')}.
\end{equation}

It is to be noted that choice of a cut-off radius is not arbitrary.

\section{Holographic unified viscous fluid description of dark energy and dark matter in mimetic matter model}

In the following section we will describe the system containing a viscous fluid in the presence of mimetic dark matter term $\lambda$.

We consider the effective inhomogeneous EOS in flat FRW space time~\cite{Nojirie,Capozziello} 
\begin{equation} 
\label{eq18} p =  \omega(\rho,t) \rho + f(\rho) -3H\zeta(H,t),
\end{equation}
where $\omega(\rho,t)$ is the thermodynamics EOS parameter and $\zeta(H,t)$ is the bulk viscosity depends on Hubble parameter $H$ and on the cosmic time $t$, we can consider the bulk viscosity is positive. 

Let us consider the unified description of the early universe and the late universe. For this we choose $f(\rho)$ in the following form \cite{Elizaldea}
\begin{equation} 
\label{eq19} \rho =  \frac{\gamma \rho^{n}}{(1+ \delta \rho^m)},
\end{equation}
where $\gamma$, $\delta$, $n$, $m$ are all free parameters.

The function $ f(\rho)$ in Eq. (18) provides description of the unified early and late universe. This form of barotropic pressure has the advantage of interpolation between different powers of density, allows for smooth transitions during the  universe evolution. It is to be noted that in this work the description part is the bulk viscosity with a constant coefficient. 

The dissipative process are described with bulk viscosity in the form \cite{Capozziello} 
\begin{equation} 
\label{eq20} \zeta(H,t) = \xi_{1}(t)(3H)^q, 
\end{equation}
where the $q$ is the parameter.

The energy conservation takes the standard form
\begin{equation} \label{eq21} 
\dot\rho + 3H(\rho + p) = 0,
\end{equation}

In the following subsections we shall discuss the three cases to elaborate the panoramic picture of the universe and in connection to explore the physical as well as cosmological features therein.

\subsection{Case I}

Consider the simple case EOS parameter $\omega = \omega_{0}$ and $\zeta(H,t) = \xi_{0}$. We also restrict to the value of $ m = \frac{1}{2}$ and $ n = m+1$. Then inhomogeneous Eq. (18) with the help of Eqs. (19) and (20) becomes 
\begin{equation} 
\label{eq22} p =  \omega_{0}\rho + \frac{\gamma \rho^{3/2}}{(1+ \delta \rho^{1/2})} -3\xi_{0}H.
\end{equation}

Then the Friedmann Eq. (8) with the mimetic matter field (13) for $\beta = V= 0$ is taken into account 
\begin{equation} 
\label{eq23} \rho  = \frac{3}{\kappa^2} H^2 - \frac{\lambda_{0}}{a^3}. 
\end{equation}

By using Eqs. (22) and (23) in the approximation of the large $\rho$ we obtain from conservation Eq. (21) 
\begin{equation} 
\label{eq24} \frac{(a\ddot a -\dot a^2)}{a^2}+ \frac{3}{2}(b_{0} +1)\frac{\dot a^2}{a^2}-\frac{\kappa^2 \lambda_{0} b_{0}}{2}\frac{1}{a^3}- \frac{3\kappa^2 \xi_{0}}{2} \left(\frac{\dot a}{a}\right)= 0,
\end{equation}
where $ b_{0}= \left(\omega_{0} +\frac{\gamma}{\delta} \right)$.

The above equation takes the from 
\begin{equation} 
\label{eq25} a\ddot a + \frac{3}{2}\left(b_{0} +\frac{1}{3}\right)\dot a^2 -\left( \frac{\kappa^2 \lambda_{0} b_{0}}{2}\right)\left(\frac{1}{a}\right) - \left(\frac{3\kappa^2 \xi_{0}}{2}\right) \dot a a = 0.
\end{equation}

In the above equation if we take $ b_{0}= - \frac{1}{3} $, then Eq. (25) reduces to 
\begin{equation} 
\label{eq26} \ddot a -\left( \frac{3\kappa^2 \xi_{0}}{2}\right) \dot a a +\left( \frac{\kappa^2 \lambda_{0}}{6}\right)\left(\frac{1}{a^2}\right) = 0. 
\end{equation}

If we consider the $\xi_{0} \rightarrow 0$ then the solution of Eq. (26) in terms of scale factor can be written as
 \begin{equation} 
\label{eq27} \left(\frac{1}{\sqrt{2C_{1}}}\right) X(a)-\left(\frac{\kappa^{2}\lambda_{0}}{\sqrt{2}\; 6(C_{1})^{3/2}}\right) ln Y(a) = t + C_{2},
\end{equation}
with 
$$X(a) = \left(\sqrt{a(C_{1}a +\frac{\kappa^2 \lambda_{0}}{6})}\right),$$
$$Y(a)= \left( \frac{\sqrt{6 C_{1}a}}{\kappa \sqrt{\lambda_{0}}}( 1+ \sqrt{1 + \frac{\kappa^2 \lambda_{0}}{6 C_{1} a}})\right),$$ 
whereas $ C_{1} \ne 0 $ and $C_{2}$ are arbitrary constants.

Let us consider the particular case when $C_{1} = C_{2} = 0 $ then solution of the Eq. (27) becomes
\begin{equation} 
\label{eq28} a(t) =\lambda_{a} t^{2/3}, 
\end{equation}
where $\lambda_{a} = (\frac{3}{2})^{2/3}(\frac{1}{3}\kappa^2\lambda_{0})^{1/3}.$

The Hubble parameter can be express as 
\begin{equation} 
\label{eq29} H(t) =\frac{2}{3t}. 
\end{equation}

From this equation it is noted that as $ t \rightarrow \infty$, the Hubble parameter $H \rightarrow 0$, and the particle horizon from Eq. (17) becomes 
\begin{equation} 
\label{eq30} L_{p} = 3t. 
\end{equation}

Nojiri and Odintsov \cite{Nojiria} obtained the expression of the Hubble parameter in term of the particle horizon and its time derivative
\begin{equation} 
\label{eq31} H = \frac{\dot L_{p} -1}{L_{p}},\; \dot H = \frac{\ddot L_{p}}{L_{p}}- \frac{\ddot L_{p}^2}{L_{p}^2} + \frac{\dot L_{p}}{L_{p}^2}.
\end{equation}

The scale factor and its time derivative in terms of the particle horizon can be written as 
\begin{equation} 
\label{eq32} a(t) =\lambda_{a} \left (\frac{\dot L_{p}}{L_{p}}\right)^{2/3},\;\;\dot a(t) =\frac{2}{3}\lambda_{a} \left (\frac{\dot L_{p}}{L_{p}}\right)^{1/3}, \;\ddot a(t) =-\frac{2}{9}\lambda_{a} \left (\frac{\ddot L_{p}}{L_{p}}\right)^{4/3}.
\end{equation}

By using Eq. (32) the energy conservation equation can be written as 
\begin{equation} 
\label{eq33} \lambda_{a} \left (\frac{2\dot L_{p}}{9L_{p}}+ \xi_{0} \kappa^2\right) - \frac{\kappa^2\lambda_{0}}{6 \lambda_{a}^{2}} \left (\frac{\dot L_{p}}{L_{p}}\right)^{1/3} =0. 
\end{equation}

The viscous fluid model containing mimetic matter is holographically described by this equation. We have so successfully used the model to apply the holographic principle. The similar outcomes previously acquired by \cite{Brevikb} within the context of general theory gravity.

%%%%%%%%%%%%%%%%%%%%%%%%%%%%%%%%%%%%%%%%%%%%%%%%%%%%%%%%%%%%%%
\begin{figure}[htbp]
\centering
\includegraphics[height=12.0cm,width=12.0cm]{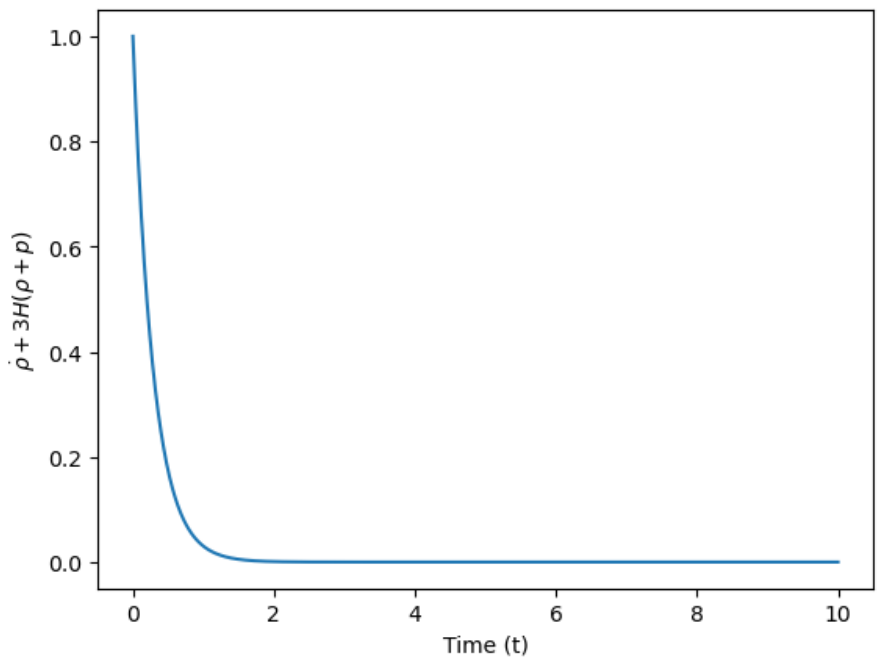}
\caption {Graphical plot of the energy conservation equation}
\label{Fig1-1}
\end{figure}

%%%%%%%%%%%%%%%%%%%%%%%%%%%%%%%%%%%%%%%%%%%%%%%%%%%%%%%%%%%%%%

%%%%%%%%%%%%%%%%%%%%%%%%%%%%%%%%%%%%%%%%%%%%%%%%%%%%%%%%%%%%%%
\begin{figure}[htbp]
\centering
\includegraphics[height=18.0cm,width=14.0cm]{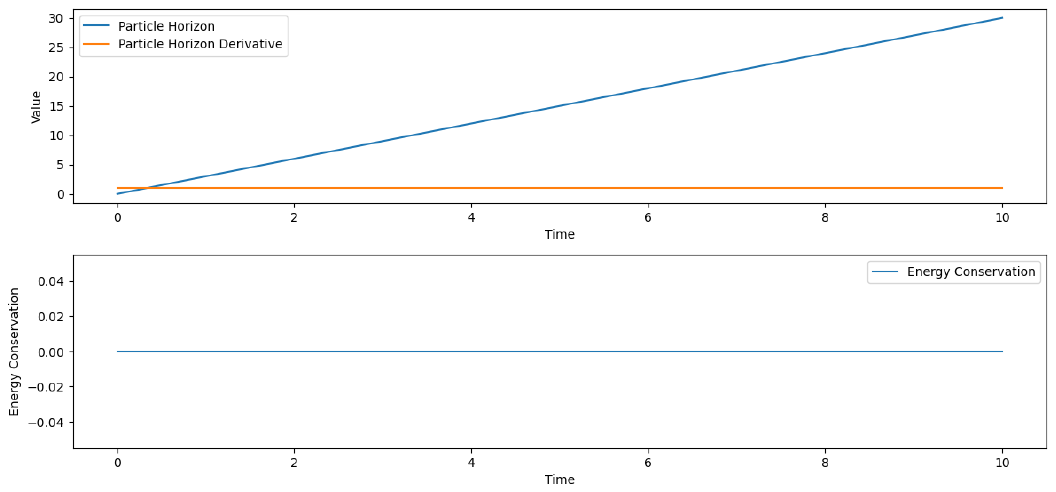}
\caption{Particle horizon and derivative vs time (upper panel) and energy conservation vs time  (lower panel) }
\end{figure}
%%%%%%%%%%%%%%%%%%%%%%%%%%%%%%%%%%%%%%%%%%%%%%%%%%%%%%%%%%%%%%

%%%%%%%%%%%%%%%%%%%%%%%%%%%%%%%%%%%%%%%%%%%%%%%%%%%%%%%%%%%%%%
\begin{figure}[thbp]
\centering
\includegraphics[height=12.0cm,width=12.0cm]{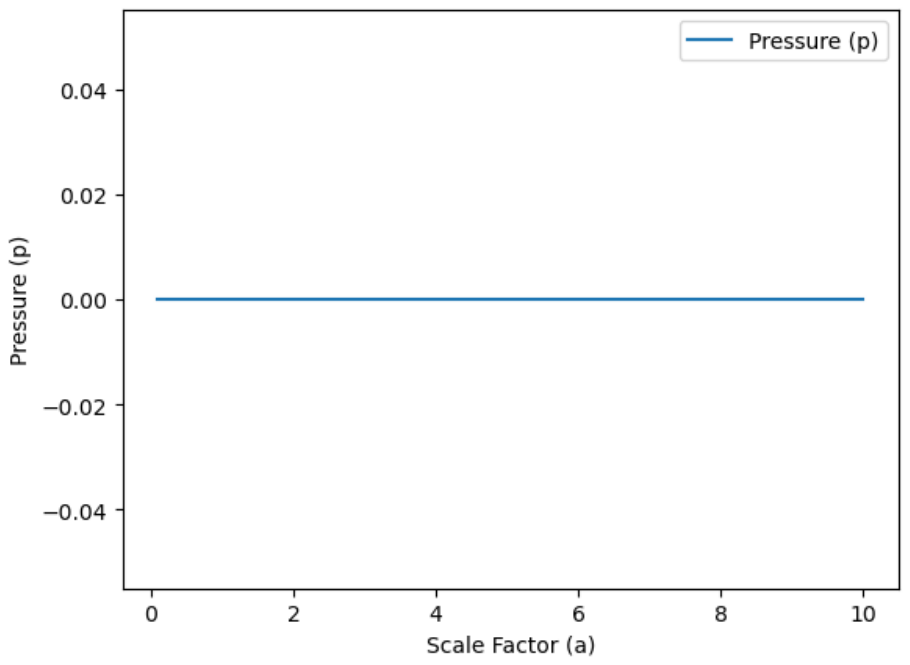}
\caption{Pressure vs scale factor }
\end{figure}
%%%%%%%%%%%%%%%%%%%%%%%%%%%%%%%%%%%%%%%%%%%%%%%%%%%%%%%%%%%%%%

%%%%%%%%%%%%%%%%%%%%%%%%%%%%%%%%%%%%%%%%%%%%%%%%%%%%%%%%%%%%%%
\begin{figure}[thbp]
\centering
\includegraphics[height=12.0cm,width=12.0cm]{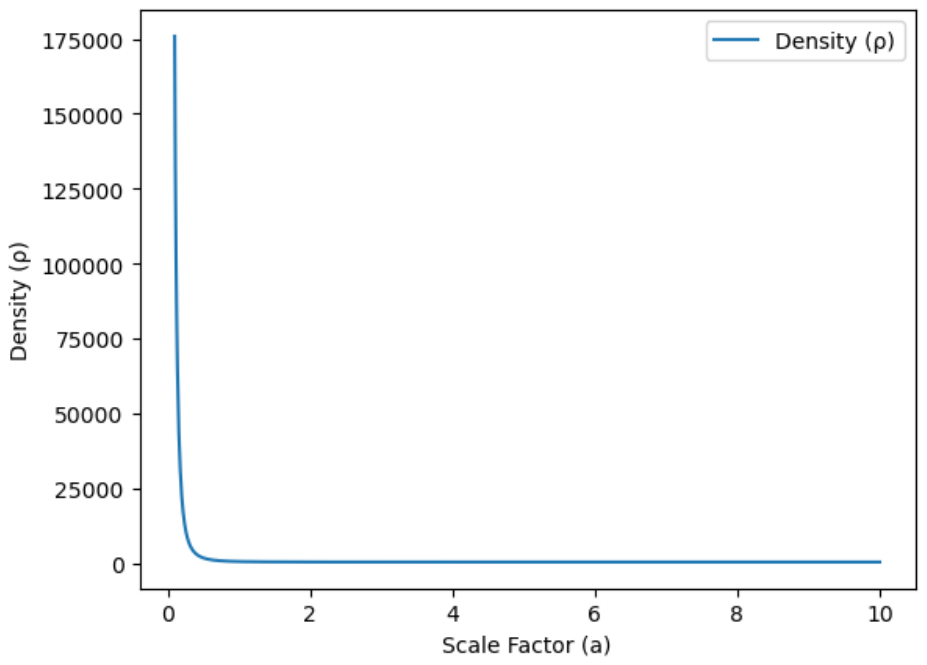}
\caption{Density vs scale factor   }
\end{figure}
%%%%%%%%%%%%%%%%%%%%%%%%%%%%%%%%%%%%%%%%%%%%%%%%%%%%%%%%%%%%%%

%%%%%%%%%%%%%%%%%%%%%%%%%%%%%%%%%%%%%%%%%%%%%%%%%%%%%%%%%%%%%%
\begin{figure}[thbp]
\centering
\includegraphics[height=12.0cm,width=12.0cm]{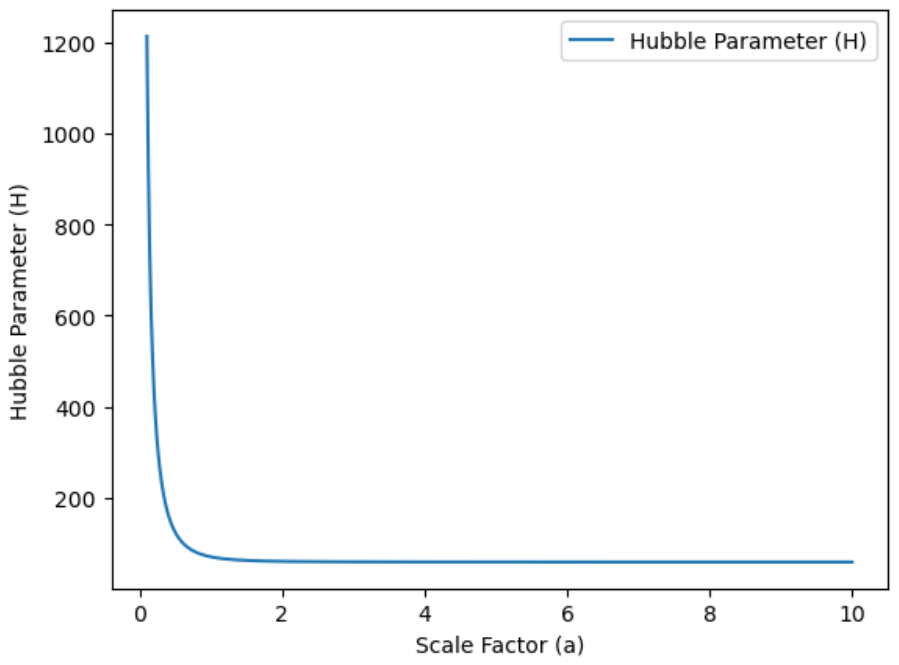}
\caption{Hubble parameter vs scale factor}
\end{figure}
%%%%%%%%%%%%%%%%%%%%%%%%%%%%%%%%%%%%%%%%%%%%%%%%%%%%%%%%%%%%%%

%%%%%%%%%%%%%%%%%%%%%%%%%%%%%%%%%%%%%%%%%%%%%%%%%%%%%%%%%%%%%%
\begin{figure}[thbp]
\centering
\includegraphics[height=12.0cm,width=12.0cm]{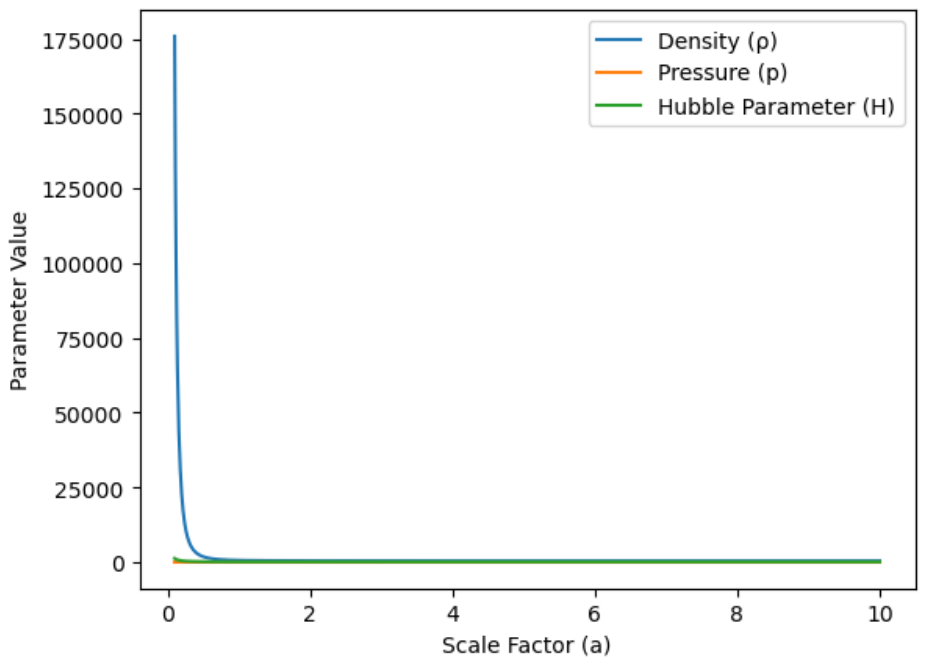}
\caption{Various parameters vs scale factor}
\end{figure}
%%%%%%%%%%%%%%%%%%%%%%%%%%%%%%%%%%%%%%%%%%%%%%%%%%%%%%%%%%%%%%

%%%%%%%%%%%%%%%%%%%%%%%%%%%%%%%%%%%%%%%%%%%%%%%%%%%%%%%%%%%%%%
\begin{figure}[thbp]
\centering
\includegraphics[height=12.0cm,width=12.0cm]{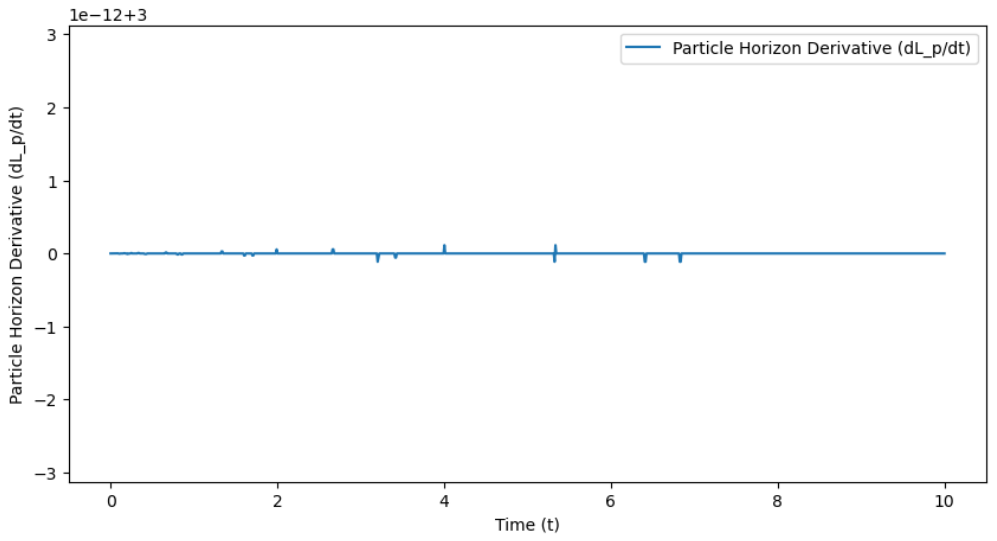}
\caption{Particle horizon derivative vs time}
\end{figure}
%%%%%%%%%%%%%%%%%%%%%%%%%%%%%%%%%%%%%%%%%%%%%%%%%%%%%%%%%%%%%%

\subsection{Case II} 

In this case we consider the EOS parameter $\omega =\omega_{0}$ and viscosity $\zeta(H,t) = 3 \alpha H$.  Hence  Eq. (18) becomes
\begin{equation} 
\label{eq34} p =  \omega_{0} \rho + \frac{\gamma \rho^{3/2}}{(1+ \delta \rho^{1/2})} -9\alpha H^2.
\end{equation}

%%%%%%%%%%%%%%%%%%%%%%%%%%%%%%%%%%%%%%%%%%%%%%%%%%%%%%%%%%%%%%
\begin{figure}[thbp]
\centering
\includegraphics[height=12.0cm,width=12.0cm]{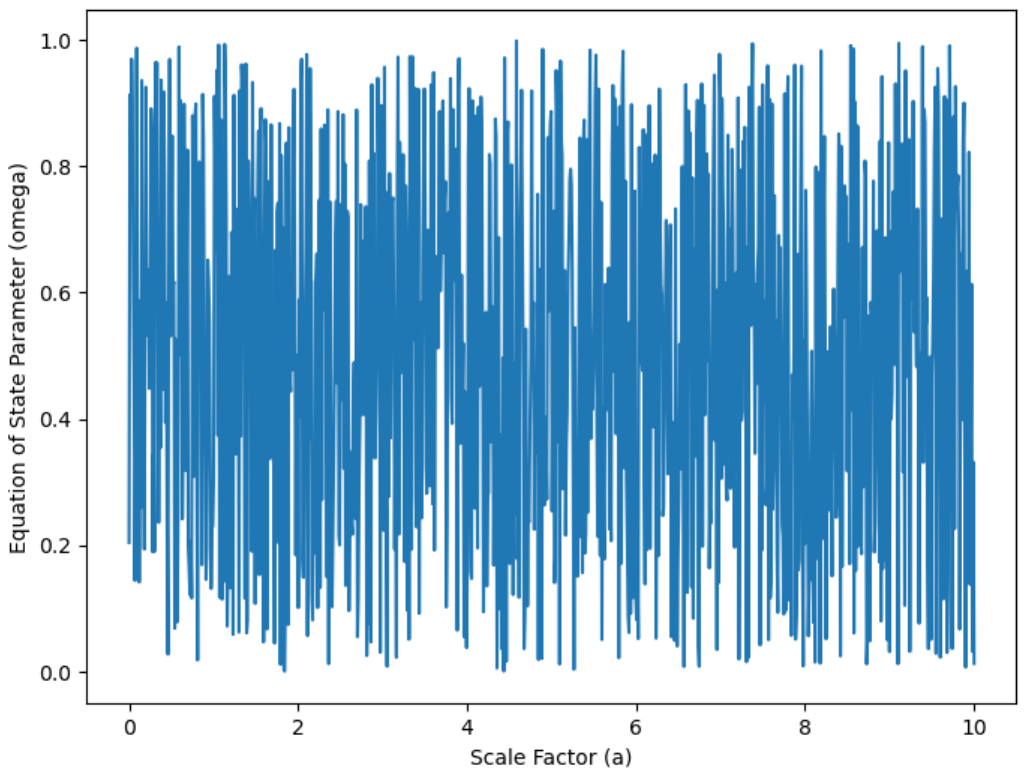}
\caption{Equation of state parameter vs scale factor}
\end{figure}
%%%%%%%%%%%%%%%%%%%%%%%%%%%%%%%%%%%%%%%%%%%%%%%%%%%%%%%%%%%%%%

%%%%%%%%%%%%%%%%%%%%%%%%%%%%%%%%%%%%%%%%%%%%%%%%%%%%%%%%%%%%%%
\begin{figure}[thbp]
\centering
\includegraphics[height=12.0cm,width=12.0cm]{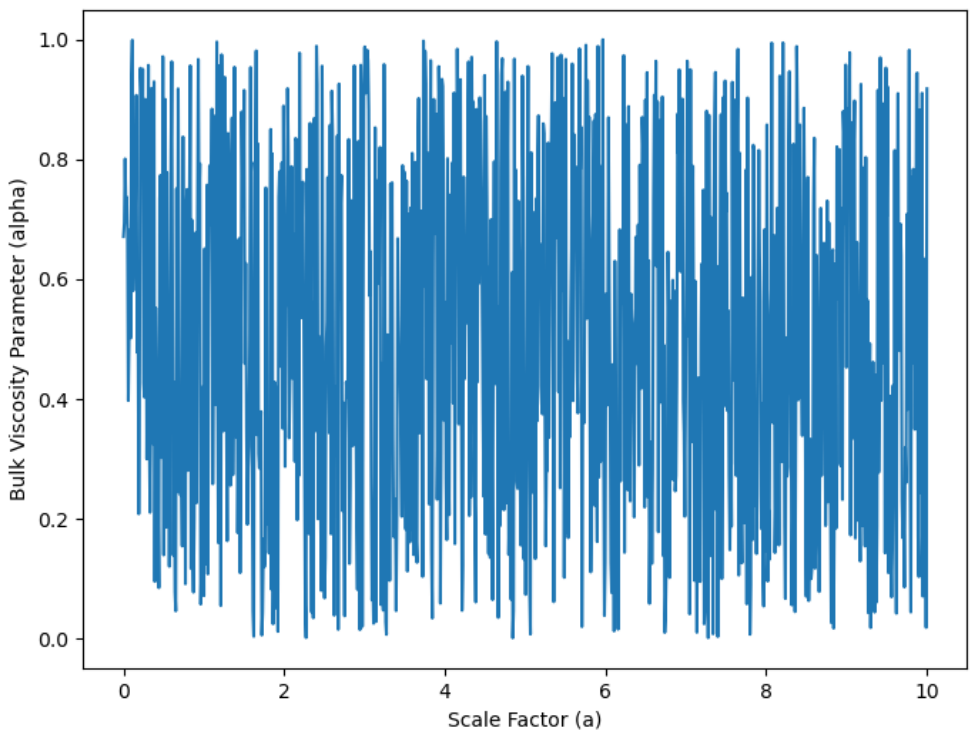}
\caption{Bulk viscosity parameter vs scale factor}
\end{figure}
%%%%%%%%%%%%%%%%%%%%%%%%%%%%%%%%%%%%%%%%%%%%%%%%%%%%%%%%%%%%%%

%%%%%%%%%%%%%%%%%%%%%%%%%%%%%%%%%%%%%%%%%%%%%%%%%%%%%%%%%%%%%%
\begin{figure}[thbp]
\centering
\includegraphics[height=12.0cm,width=12.0cm]{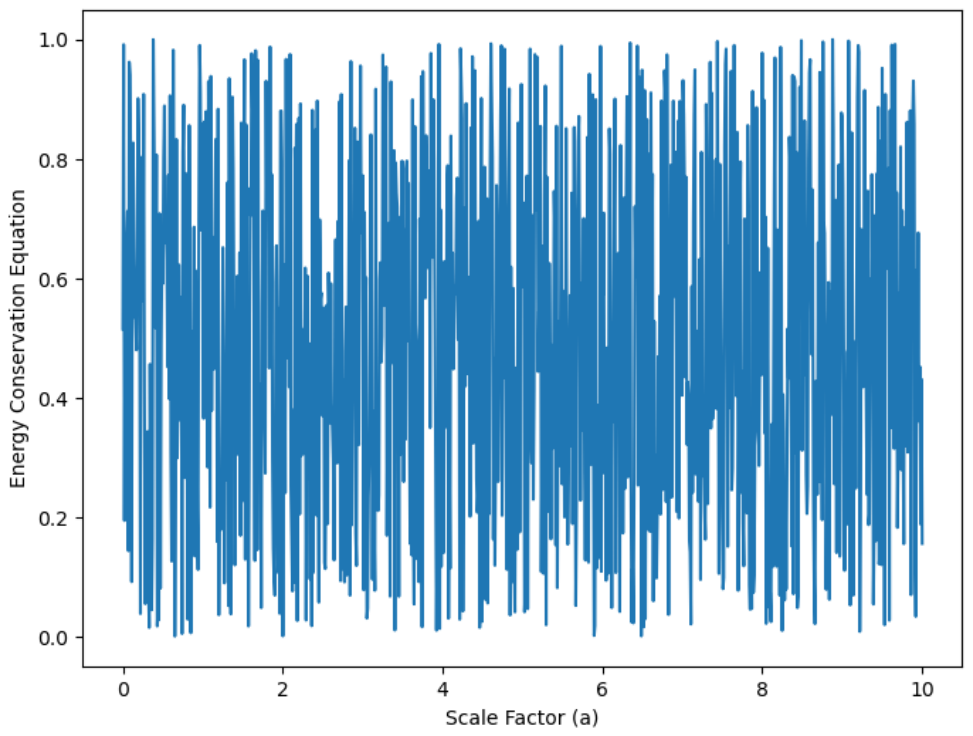}
\caption{Energy conservation equation vs scale factor}
\end{figure}
%%%%%%%%%%%%%%%%%%%%%%%%%%%%%%%%%%%%%%%%%%%%%%%%%%%%%%%%%%%%%%

With the conservation equation (21) in the approximation of large energy density $\rho$, we obtain the differential equation in terms of the scale factor $a(t)$ as follows by applying Eqs. (23) and (34)
\begin{equation} 
\label{eq35} a\ddot a + \frac{1}{2}(3\eta - 9\alpha \kappa^2 +1 )\dot a^2 -\left(\frac{\kappa^2 \lambda_{0} \eta}{2}\right) \left(\frac{1}{a}\right)=0, 
\end{equation}
where $\eta = \omega_{0} +\frac{\gamma}{\delta}$.

When $\eta = 0$, we can look at the simplified version of Eq. (35) 
\begin{equation} 
\label{eq36} a\ddot a + \frac{1}{2}(1 - 9\alpha \kappa^2)\dot a^2 = 0, 
\end{equation}
after simplifying above equation we get 
\begin{equation} 
\label{eq37} a(t) = C_{1} e^{C_{2}t} \;\; \Rightarrow H = C_{2},
\end{equation}
where $C_{1} $ and $C_{2}$ are arbitrary constants and $\frac{1}{2}(1 - 9\alpha \kappa^2)\ne 0.$ If we take $\frac{1}{2}(1 - 9\alpha \kappa^2) =-1$,  then viscous parameter is  $\alpha = \frac{1}{3}\kappa^2$. 

Similarly the particle horizon 
 \begin{equation} 
\label{eq38} L_{p} = \frac{1}{C_{1}}\left( e^{(C_{1}t -1)}\right). 
\end{equation}

Then the holographic representation of the equation of motion Eq. (36) is 
\begin{equation} 
\label{eq39}   (1+ C_{1} L_{p})\ddot L_{P} +\frac{C_{1}}{2}(1 - 9\alpha \kappa^2)\dot L_{p}^2 =0.
\end{equation}

Eq. (39) rewrites the energy conservation formula using the holographic principle.

\subsection{Case III}

In this case we consider the EOS parameter $\omega=\omega_{eff} =-1$ and viscosity $\zeta(H,t) = \xi_{0}$. We can also define the effective energy density, effective pressure and effective EOS parameter of the universe as 
\begin{equation}
\label{eq40} \rho_{eff} = \rho +\lambda + V,
\end{equation}

\begin{equation}
\label{eq41} p_{eff} = p-V,
\end{equation}

\begin{equation}
\label{eq42} \omega_{eff} =  \frac{p_{eff}}{\rho_{eff}}= \frac{p -V}{\rho +\lambda + V}\;\;.
\end{equation}

%%%%%%%%%%%%%%%%%%%%%%%%%%%%%%%%%%%%%%%%%%%%%%%%%%%%%%%%%%%%%%
\begin{figure}[thbp]
\centering
\includegraphics[height=14.0cm,width=16.0cm]{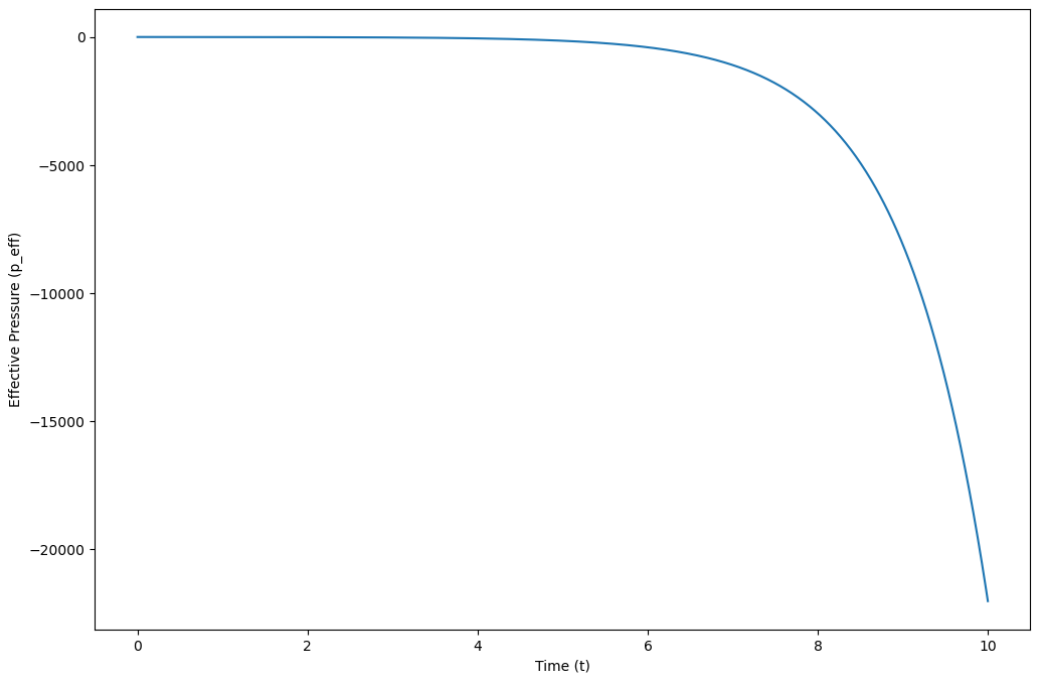}
\caption{Effective pressure vs time}
\end{figure}
%%%%%%%%%%%%%%%%%%%%%%%%%%%%%%%%%%%%%%%%%%%%%%%%%%%%%%%%%%%%%%

%%%%%%%%%%%%%%%%%%%%%%%%%%%%%%%%%%%%%%%%%%%%%%%%%%%%%%%%%%%%%%
\begin{figure}[thbp]
\centering
\includegraphics[height=14.0cm,width=16.0cm]{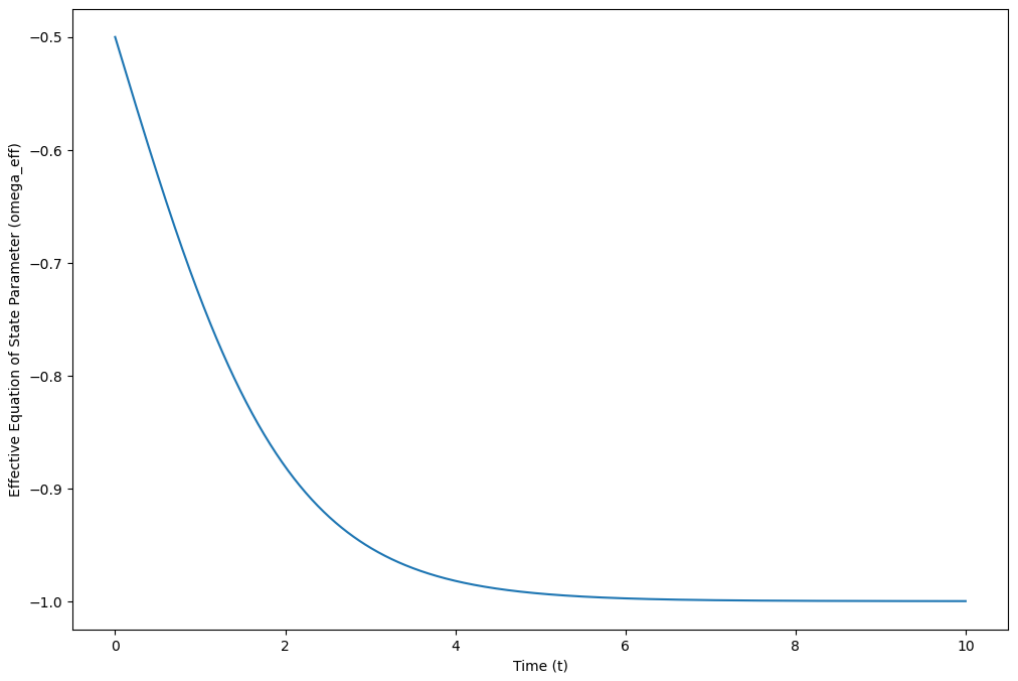}
\caption{Effective equation of state parameter vs time}
\end{figure}
%%%%%%%%%%%%%%%%%%%%%%%%%%%%%%%%%%%%%%%%%%%%%%%%%%%%%%%%%%%%%%

%%%%%%%%%%%%%%%%%%%%%%%%%%%%%%%%%%%%%%%%%%%%%%%%%%%%%%%%%%%%%%
\begin{figure}[thbp]
\centering
\includegraphics[height=4.0cm,width=10.0cm]{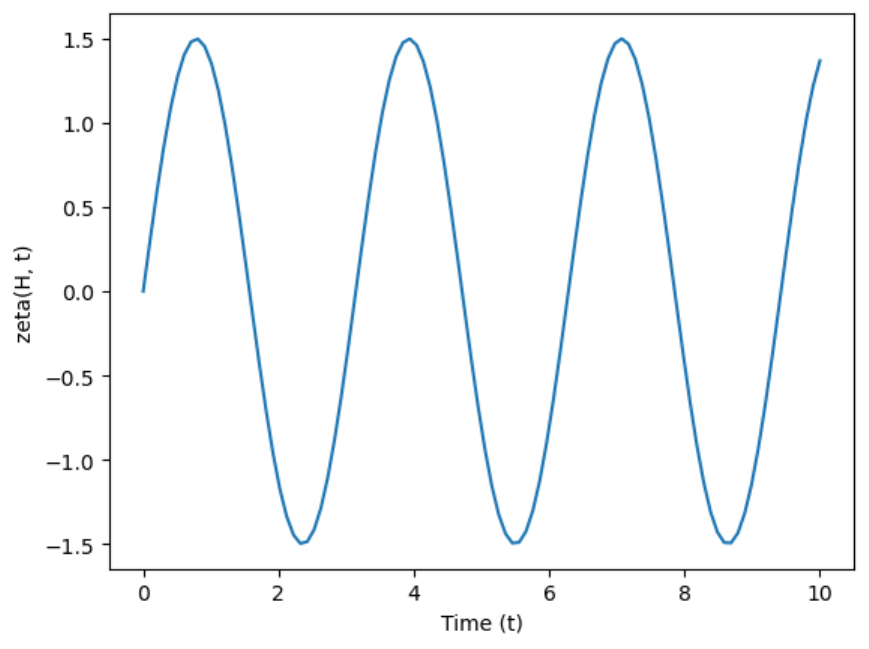}
\caption{viscosity $\zeta(H,t)$ vs time}
\end{figure}
%%%%%%%%%%%%%%%%%%%%%%%%%%%%%%%%%%%%%%%%%%%%%%%%%%%%%%%%%%%%%%

With this definition the Friedmann Eqs. (8) and (9) can be written as 
\begin{equation}
\label{eq43} 3H^2  = \kappa^2\rho_{eff},
\end{equation}

\begin{equation}
\label{eq44} 2\dot {H} + 3 H^2 = -\kappa^2 p_{eff}.
\end{equation}

From the Eqs. (43) and (44) we can arrive to the continuity equation 
\begin{equation} 
\label{eq45} \dot\rho_{eff}  + 3H(\rho_{eff}  + p_{eff} ) = 0. 
\end{equation}

Also Eq. (18) can be written as 
\begin{equation} 
\label{eq46} p_{eff} =   \omega_{eff}\; \rho_{eff} + f(\rho_{eff}) - 3 H \xi_{0},
\end{equation}

By using Eq. (19) we get 	 
\begin{equation} 
\label{eq47} p_{eff} =  - \rho_{eff} + \frac{\gamma \rho_{eff}^{n}}{1+\delta \rho_{eff}^{m} } - 3 H \xi_{0}.
\end{equation}

By using Eqs. (47) and (45), we obtain 
\begin{equation} 
\label{eq48} \dot\rho_{eff}  + 3H\left(\frac{\gamma \rho_{eff}^{n}}{1+\delta \rho_{eff}^{m} } - 3 H \xi_{0} \right) = 0. 
\end{equation}

If we restrict ourselves to $n > 0$ and $ m > 0$, then provided $ n-m  = \frac{1}{2}$  for large $\rho_{eff}$, the above equation asymptotically becomes
\begin{equation} 
\label{eq49} \dot\rho_{eff} = \bar \lambda_{0} \rho_{eff},  
\end{equation}
where $\bar \lambda_{0} = (\sqrt{3}\kappa \gamma \delta^{-1}) ( \bar\xi_{0} -1)$ and $\bar\xi_{0} = \sqrt{3}\;\kappa \xi_{0}\delta \gamma^{-1}$. 

The solution of Eq. (49) is 
\begin{equation} 
\label{eq50} \rho_{eff} = \bar \rho_{0} e^{\bar \lambda_{0}t},  
\end{equation}
where $\bar\rho_{0}$ is arbitrary constant.

Accordingly the Hubble parameter takes the from 
\begin{equation} 
\label{eq51} H = \left(\frac{\kappa \bar \rho_{0}}{\sqrt{3}}\right) e^{ \frac{\bar \lambda_{0}}{2}t}.  
\end{equation}

From Eq. (51), it is observed that as $t\rightarrow 0$, the Hubble parameter becomes constant, $H(t) \rightarrow \left(\frac{\kappa \bar \rho_{0}}{\sqrt{3}}\right).$ This case can be identified with the inflation. Hence, the scale factor is given by 
\begin{equation} 
\label{eq52} a = a_{0}exp \left(\lambda_{1} e^{\frac{\bar \lambda_{0}}{2} t} \right),
\end{equation}
where $\lambda_{1} = \frac{2\kappa \rho_{0}}{\sqrt{3} \bar \lambda_{0}}.$

We can calculate the particle horizon $L_{p}$ as 
\begin{equation} 
\label{eq53} 
L_{p} =a(t) \int_{0}^{t}\frac{dt'}{a(t')} =\frac{2}{\bar\lambda_{0}}exp \left(\lambda_{1} e^{\frac{\bar \lambda_{0}}{2} t} \right)\left[ Ei(-\lambda_{1} e^{\frac{\bar \lambda_{0}}{2} t})- Ei(-\lambda_{1})\right],
\end{equation}
where $Ei(x)$ is the integral exponential function.

From Eq. (51) after  differentiation with respect to $t$ we get
\begin{equation} 
\label{eq54} \dot H = \frac{\bar\lambda_{0}}{2}H.
\end{equation}

By using Eq. (31) the energy conservation equation becomes
 \begin{equation} 
\label{eq55} 
\frac{\ddot L_{p}}{L_{p}}- \frac{\ddot L_{p}^2}{L_{p}^2} + \frac{\dot L_{p}}{L_{p}^2}  =\frac{ \bar\lambda_{0}}{2}\left (\frac{\dot L_{p} -1}{L_{p}}\right).
\end{equation}

Thus we have successfully applied the holographic principle in the framework of mimetic gravity.

\section{Conclusion}

The essence of the mimetic dark matter concept lies in a constraint requiring the gradient to be a unit time-like vector, distinguishing it from general relativity. This theory employs a singular disformal transformation to isolate the conformal degree of freedom of gravity in a covariant manner, leading to altered dynamics and the emergence of an effective dark matter component on cosmological scales \cite{Jibitesh}.

Numerous previous studies have investigated background cosmological solutions within mimetic gravity and found that, with a suitable choice of potential for the mimetic field, attractive cosmological solutions can be attained, mimicking expansion histories consistent with observational data without the need for additional dark matter or dark energy fluids \cite{Jibitesh}.

This study delved into the holographic unified model of the early and late universe, alongside the cosmic evolution of a mimetic matter model in a homogeneous and isotropic, particularly flat FRW metric, from a holographic standpoint. Assuming the presence of a mimetic dark matter model, specifically the field mimics \cite{Bita}, we expressed the energy conservation equation in holographic language, presuming the universe to be filled with viscous dark fluid.

Utilizing the cut-off model proposed by Nojiri and Odintsov \cite{Nojiria,Nojirib}, we established in mimetic gravity the equivalence between viscous fluid cosmology and holographic fluid cosmology. We scrutinized the inhomogeneous equation of state for viscous dark fluid in the presence of mimetic matter, considering straightforward examples for the equation of state parameters $\omega = \omega_{0}$ and $\omega = -1$. To achieve this, we determined the particle horizon $L_{P}$ and the infrared radius $L_{IR}$. We explored the holographic principle for a cosmological model where the bulk viscosity $\zeta(H,t)$ varies. To derive the energy conservation equation for each scenario, we determined the infrared radius, which takes the form of a particle horizon, allowing for a unified description of the early and late universe in cases when the general equation of state includes bulk viscosity.

The role of mimetic potential, or $V \propto \lambda$, in dark energy is believed to be proportional to the mimetic dark matter density. By introducing an interaction between dark matter and dark energy, we established $\lambda \approx a^{-3(\beta +1)}$, leading to three distinct scenarios.

In Case I, the Hubble parameter becomes constant, $H \rightarrow 0$, resulting in late-time accelerating expansion for the late universe as $t \rightarrow \infty$.

For the second example from Eq. (37), the Hubble parameter tends to a constant, $H \rightarrow C_{2}$, corresponding to inflation for both the early and late universe, i.e., $t \rightarrow 0$ and $t \rightarrow \infty$.

In Case III, based on Eq. (51), the Hubble parameter $H(t) \rightarrow \frac{\kappa \bar \rho_{0}}{\sqrt{3}}$ becomes constant as $ t \rightarrow 0$, comparable to inflation. We expounded on the holographic picture using a viscous fluid \cite{Brevikb}, although in the inflationary scenario, the contribution of bulk viscosity is typically negligible and interacts solely with the universe's evolution. Consequently, we demonstrated the equivalence between the holographic model based on the choice of the infrared radius and the description of the unified model of the early and late universe in the presence of viscous fluid within the framework of mimetic gravity.

\section*{Data Availability}             
No new data were generated in support of this research.

\section*{Conflict of Interest} 
The authors declare no conflict of interest.

\section*{Acknowledgements}

The necessary literature was obtained during a visit to the Inter University Center for Astronomy and Astrophysics (IUCAA) in Pune, for which GSK expresses gratitude. SR gratefully acknowledges the support from IUCAA, Pune, India, provided through the Visiting Research Associateship Programme, and the facilities available at ICARD, Pune, at CCASS, GLA University, Mathura. We also appreciate the guidance and assistance of Professor Farook Rahaman.

\end{document}